\documentclass[12pt,a4paper]{article}  
\usepackage[dvips]{graphicx}  
\begin{document} 
\begin{titlepage} 
\begin{flushright} 
LNF--01/021(P)\\ 
UAB--FT--515\\ 
hep-ph/0105179\\ 
May 2001 
\end{flushright} 
\vspace*{1.6cm} 
 
\begin{center} 
{\Large\bf  
Scalar $\sigma$ meson effects in $\rho$ and $\omega$ decays\\ 
into $\pi^0\pi^0\gamma$}\\ 
\vspace*{0.8cm} 
 
A.~Bramon$^1$, R.~Escribano$^2$, J.L.~Lucio M.$^3$ and M.~Napsuciale$^3$\\ 
\vspace*{0.2cm} 
 
{\footnotesize\it 
$^1$Departament de F\'{\i}sica, 
Universitat Aut\`onoma de Barcelona, E-08193 Bellaterra (Barcelona), Spain\\ 
$^2$INFN-Laboratori Nazionali di Frascati, P.O.~Box 13, I-00044 Frascati, Italy\\ 
$^3$Instituto de F\'{\i}sica, Universidad de Guanajuato, 
Lomas del Bosque \#103, Lomas del Campestre, 37150 Le\'on, Guanajuato, Mexico} 
\end{center} 
\vspace*{1.0cm} 
 
\begin{abstract} 
The complementarity between Chiral Perturbation Theory and the Linear Sigma Model  
in the scalar channel is exploited to study $\pi^0\pi^0$ production in $\rho$ and  
$\omega$ radiative decays, where the effects of a low mass scalar resonance   
$\sigma(500)$ should manifest.  
The recently reported data on $\rho\rightarrow\pi^0\pi^0\gamma$ seem to require the 
contribution of a low mass and moderately narrow $\sigma(500)$. 
The properties of this controversial state could be fixed by improving the accuracy 
of these measurements. 
Data on $\omega\rightarrow\pi^0\pi^0\gamma$ can also be accommodated in our framework, 
but are much less sensitive to the $\sigma(500)$ properties.    
\end{abstract} 
\end{titlepage} 
\section{Introduction} 
 
Radiative decays of vector mesons have gained renewed interest as a useful tool  
to improve our insight into the complicated dynamics governing meson physics  
in the 1 GeV energy region.  
Particularly interesting are those decays proceeding by the exchange of scalar  
resonances because of the enigmatic nature of these states and the poor knowledge  
on their properties. 
In the case of the $\sigma$ meson  
---a broad and controversial scalar state with a mass peaked somewhere in the  
500 MeV region--- 
the situation is even more dramatic: 
the issue under discussion along the years has been the existence or not of such a  
state. 
 
The SND Collaboration has reported very recently the first measurement of the  
$\rho\rightarrow\pi^0\pi^0\gamma$ decay.  
For the branching ratio, they obtain \cite{Achasov:2000zr} 
\begin{equation} 
\label{SND}  
B(\rho\rightarrow\pi^0\pi^0\gamma)=(4.8^{+3.4}_{-1.8}\pm 0.2)\times 10^{-5}\   
\end{equation} 
and therefore $\Gamma(\rho\rightarrow\pi^0\pi^0\gamma)=(7.2^{+5.1}_{-2.7})$ keV. 
For the analogous $\omega$ radiative decay, the GAMS Collaboration reported some  
years ago the branching ratio \cite{Alde:1994kf} 
\begin{equation} 
\label{GAMS}  
B(\omega\rightarrow\pi^0\pi^0\gamma)=(7.2\pm 2.5)\times 10^{-5}\ ,  
\end{equation} 
which implies $\Gamma(\omega\rightarrow\pi^0\pi^0\gamma)=(608 \pm 211)$ eV.  
The result in Eq.~(\ref{GAMS}) has been confirmed by the more recent but  
less accurate measurement by the SND Collaboration  
$B(\omega\rightarrow\pi^0\pi^0\gamma)=(7.8\pm 2.7\pm 2.0)\times 10^{-5}$ 
\cite{Achasov:2000zr}.  
Since $m_{\rho}\simeq m_{\omega}\simeq 780$ MeV, both processes contain valuable  
information on the scalar channel of the $\pi^0\pi^0$ system in the range of  
masses where the $\sigma(500)$ resonance effects are expected to manifest.   
These and other radiative vector meson decays will be hopefully investigated  
at the Frascati $\phi$-factory DA$\Phi$NE very soon \cite{daphne95:franzini}.  
 
On the theoretical side, the $V\rightarrow P^0 P^0\gamma$ decays have been  
considered by a number of authors \cite{LucioMartinez:1990uw}--\cite{Gokalp:2000xy}. 
Early calculations of the vector meson dominance (VMD) amplitude for these processes, 
{\it i.e.}~the contributions proceeding through the decay chain  
$V\rightarrow  P^0 V^\prime\rightarrow P^0 P^0\gamma$, were summarized in  
Ref.~\cite{Bramon:1992kr}. 
In particular, the widths and branching ratios predicted by VMD,      
$\Gamma(\rho\rightarrow\pi^0\pi^0\gamma)_{\rm VMD}=1.62$ keV,   
$B(\rho\rightarrow\pi^0\pi^0\gamma)_{\rm VMD}=1.1 \times 10^{-5}$,     
$\Gamma(\omega\rightarrow\pi^0\pi^0\gamma)_{\rm VMD}=235$ eV and  
$B(\omega\rightarrow\pi^0\pi^0\gamma)_{\rm VMD}=2.8 \times 10^{-5}$,   
were found to be substantially smaller than the experimental results quoted  
in Eqs.~(\ref{SND},\ref{GAMS}).  
 
The possibility of an enhancement in the first branching ratio through the 
$\rho\rightarrow\pi^+\pi^-\gamma\rightarrow\pi^0\pi^0\gamma$ mechanism  
was pointed out in Ref.~\cite{Bramon:1992kr} and further discussed   
in Ref.~\cite{Bramon:1992ki}   
in a Chiral Perturbation Theory (ChPT) context  
enlarged to include on-shell vector mesons.  
This formalism gives well defined predictions for the various  
$V\rightarrow P^0 P^0\gamma$ decays in terms of $P^+P^-\rightarrow P^0 P^0$ 
rescattering amplitudes, which are easily calculated in strict ChPT, 
and a loop integral over the intermediate $P^+P^-$ pair. 
In this approach, the $\rho\rightarrow\pi^0\pi^0\gamma$ decay is dominated by  
pion loops leading to $\Gamma(\rho\rightarrow\pi^0\pi^0\gamma)_{\chi}=1.42$ keV,   
while kaon loop contributions are three orders of magnitude smaller.  
The interference between this pion loop contribution and the previous  
VMD amplitude turns out to be constructive leading globally to   
$\Gamma(\rho\rightarrow\pi^0\pi^0\gamma)_{{\rm VMD}+\chi}=3.88$ keV and 
$B(\rho\rightarrow\pi^0\pi^0\gamma)_{{\rm VMD}+\chi}=2.6\times 10^{-5}$  
\cite{Bramon:1992ki},  
which are still small compared to the experimental result in Eq.~(\ref{SND}). 
 
The analysis of the $\omega\rightarrow\pi^0\pi^0\gamma$ decay is more involved.   
Ignoring $\rho$-$\omega$ mixing, pion loops are forbidden because of $G$-parity 
and kaon loops should now account for the whole chiral loop contribution to this process.  
However, this contribution is also small because of the relatively large kaon mass.  
As a result, the $\omega\rightarrow\pi^0\pi^0\gamma$ transition is then dominated  
by the VMD contribution that predicts  
$\Gamma(\omega\rightarrow\pi^0\pi^0\gamma)=235$ eV and 
$B(\omega\rightarrow\pi^0\pi^0\gamma)=2.8\times 10^{-5}$ \cite{Bramon:1992ki}, 
a value which is nearly two standard deviations below the experimental result  
in Eq.~(\ref{GAMS}).  
Recently, this process has been reanalyzed by Guetta and Singer \cite{Guetta:2001ra}  
who have explored the possibility of $\rho$-$\omega$ mixing effects  
bringing into the game the pion loop and vector meson contributions of the  
previously discussed $\rho\rightarrow\pi^0\pi^0\gamma$ process. 
Their final prediction is then  
$\Gamma(\omega\rightarrow\pi^0\pi^0\gamma)=(390\pm 96)$ eV and 
$B(\omega\rightarrow\pi^0\pi^0\gamma)=(4.6\pm 1.1)\times 10^{-5}$. 
 
Since the theoretical predictions for the decays  
$\rho,\omega\rightarrow\pi^0\pi^0\gamma$ are still far from the experimental values  
quoted in Eqs.~(\ref{SND},\ref{GAMS}) additional contributions are certainly required.  
The most natural candidates for closing the gap between theory and experiment  
are the contributions coming from the exchange of scalar resonances such as  
the well established $f_0(980)$ and the more controversial $\sigma(500)$  
(or $f_0$(400--1200)) mesons \cite{Groom:2000in}.  
The $\rho,\omega\rightarrow\pi^0\pi^0\gamma$ decays are thus an excellent place  
to study the properties of the elusive $\sigma(500)$ meson, 
which is supposed to couple strongly to low mass pion pairs,  
while the corresponding $\phi\rightarrow\pi^0\pi^0\gamma$ decay is more suitable 
for fixing the properties of the heavier $f_0(980)$ meson. 
 
A first analysis in this direction was done in Ref.~\cite{Marco:1999df} where the  
$\rho\rightarrow\pi^0\pi^0\gamma$ decay was considered in the framework of the  
Unitarized Chiral Perturbation Theory (U$\chi$PT).  
By a unitary resummation of the pion loop effects, these authors obtained 
$B(\rho\rightarrow\pi^0\pi^0\gamma)_{\rm U\chi PT}=1.4\times 10^{-5}$ 
and noted in passing that this result could be interpreted as a manifestation of  
the mechanism $\rho\rightarrow\sigma\gamma\rightarrow\pi^0\pi^0\gamma$.  
A later attempt describing scalar resonance effects in this process appeared  
more recently in Ref.~\cite{Gokalp:2000ir}.  
An exceedingly large width for the scalar dominated  
$\rho\rightarrow\pi^0\pi^0\gamma$ decay process, 
$\Gamma(\rho\rightarrow\pi^0\pi^0\gamma)=$289 keV,  
is obtained using a $\sigma$ pole model \cite{Gokalp:2000ir}.  
This unrealistic result is a consequence of using a large and constant 
$\rho\rightarrow\sigma\gamma$ amplitude \cite{Gokalp:2000xe} quite different from  
that predicted by the Linear Sigma Model (L$\sigma$M) where it turns out to be a  
momentum dependent amplitude induced at the one loop level. 
In the L$\sigma$M approach, the Goldstone boson nature of the pions and their  
derivative couplings are a consequence of the cancellations between the pointlike 
four-pion vertex and the $\sigma$ exchange contributions (see below). 
This latter cancellations do not occur in the treatment of Ref.~\cite{Gokalp:2000ir}. 
 
The purpose of this note is to study the effects of the low mass scalar states in the  
$\rho,\omega\rightarrow\pi^0\pi^0\gamma$ decays following the ChPT inspired context 
introduced in Ref.~\cite{Bramon:2000vu} to account similarly for the 
$a_0(980)$ exchange contributions to $\phi\rightarrow\pi^0\eta\gamma$.  
In this context one takes advantage of the common origin of ChPT and the L$\sigma$M 
to improve the chiral loop predictions for $V\rightarrow P^0 P^0\gamma$ exploiting  
the complementarity of both approaches for these specific processes. 
As a result, simple analytic amplitudes,  
${\cal A}(\rho,\omega\rightarrow\pi^0\pi^0\gamma )_{\mbox{\scriptsize L$\sigma$M}}$,  
will be obtained which include the effects of the scalar meson poles and also show  
the appropriate behaviour expected from ChPT at low dipion invariant masses.  
Unlike the $\phi\rightarrow\pi^0\eta\gamma$ decay studied in Ref.~\cite{Bramon:2000vu}, 
there also exist important contributions to $\rho,\omega\rightarrow\pi^0\pi^0\gamma$  
coming from the previously mentioned vector meson exchanges. 
These VMD amplitudes,  
${\cal A}(\rho,\omega\rightarrow\pi^0\pi^0\gamma )_{\mbox{\scriptsize VMD}}$,  
are well known and scarcely interesting but have to be added to   
${\cal A}(\rho,\omega\rightarrow\pi^0\pi^0\gamma )_{\mbox{\scriptsize L$\sigma$M}}$,  
{\it i.e.}~to the relevant amplitudes containing the scalar meson effects,  
in order to compare with available and forthcoming data. 
We will conclude that data on the $\rho\rightarrow\pi^0\pi^0\gamma$ channel with a  
precision around 10\% would be sufficient to decisively improve our knowledge on the  
scalar states and, in particular, on the controversial low mass $\sigma$ meson.   
  
\section{Chiral loop contributions to $\rho\rightarrow\pi^0\pi^0\gamma$} 
\label{sectChPT} 
 
The vector meson initiated $V\rightarrow P^0 P^0\gamma$ decays cannot be treated in  
strict Chiral Perturbation Theory (ChPT).  
This theory has to be extended to incorporate on-shell vector meson fields. 
At lowest order, this may be easily achieved by means of the ${\cal O}(p^2)$  
ChPT Lagrangian 
\begin{equation} 
\label{ChPTlag} 
{\cal L}_2= 
\frac{f^2}{4}\langle D_\mu U^\dagger D^\mu U+M(U+U^\dagger )\rangle\ , 
\end{equation} 
where $U=\exp(i\sqrt{2}P/f)$ with $P$ being the usual pseudoscalar nonet matrix,  
and, at this order, $f=f_\pi=92.4$ MeV and  
$M=\mbox{diag}(m_\pi^2, m_\pi^2, 2m_K^2-m_\pi^2)$. 
The covariant derivative, now enlarged to include vector mesons, is defined as  
$D_\mu U=\partial_\mu U -i e A_\mu [Q,U] - i g [V_\mu,U]$ with  
$Q=\mbox{diag}(2/3, -1/3, -1/3)$ being the quark charge matrix and $V_\mu$  
the additional matrix containing the nonet of ideally mixed vector meson fields.  
We follow the conventional normalization for the vector nonet matrix such that the  
diagonal elements are $(\rho^0+\omega)/\sqrt{2}, (-\rho^0+\omega)/\sqrt{2}$ and $\phi$. 
 
We start considering the $\rho\rightarrow\pi^0\pi^0\gamma$ amplitude.  
There is no tree-level contribution from the Lagrangian (\ref{ChPTlag}) to this  
amplitude and at the one-loop level one needs to compute the set of diagrams shown in  
Ref.~\cite{Bramon:1992ki}.  
We do not take into account kaon loop contributions here since they were shown to be  
negligible as compared to those from pion loops \cite{Bramon:1992ki}. 
A straightforward calculation leads to the following {\it finite} amplitude for  
$\rho(q^\ast,\epsilon^\ast)\rightarrow\pi^0(p)\pi^0(p^\prime)\gamma(q,\epsilon)$ 
(see Ref.~\cite{Bramon:1992ki} for further details): 
\begin{equation} 
\label{ArhoChPT} 
{\cal A}(\rho\rightarrow\pi^0\pi^0\gamma)_\chi=  
\frac{-eg}{\sqrt{2}\pi^2 m^2_{\pi^+}}\,\{a\}\,L(m^2_{\pi^0\pi^0})\times 
{\cal A}(\pi^+\pi^-\rightarrow\pi^0\pi^0)_\chi\ , 
\end{equation}  
where 
$\{a\}=(\epsilon^\ast\cdot\epsilon)\,(q^\ast\cdot q)- 
       (\epsilon^\ast\cdot q)\,(\epsilon\cdot q)$ 
makes the amplitude Lorentz- and gauge-invariant,  
$m^2_{\pi^0\pi^0}\equiv s\equiv (p+p^\prime)^2=(q^\ast -q)^2$ is the invariant mass of  
the final pseudoscalar system and $L(m^2_{\pi^0\pi^0})$ is the loop integral function  
defined as 
\begin{equation} 
\label{L} 
\begin{array}{rl} 
L(m^2_{\pi^0\pi^0}) &=\  
\frac{1}{2(a-b)}- 
\frac{2}{(a-b)^2}\left[f\left(\frac{1}{b}\right)-f\left(\frac{1}{a}\right)\right]\\[2ex] 
&+\ \frac{a}{(a-b)^2}\left[g\left(\frac{1}{b}\right)-g\left(\frac{1}{a}\right)\right]\ . 
\end{array} 
\end{equation} 
Here 
\begin{equation} 
\label{f&g} 
\begin{array}{l} 
f(z)=\left\{ 
\begin{array}{ll} 
-\left[\arcsin\left(\frac{1}{2\sqrt{z}}\right)\right]^2 & z>\frac{1}{4}\\[1ex] 
\frac{1}{4}\left(\log\frac{\eta_+}{\eta_-}-i\pi\right)^2 & z<\frac{1}{4} 
\end{array} 
\right.\\[5ex] 
g(z)=\left\{ 
\begin{array}{ll} 
\sqrt{4z-1}\arcsin\left(\frac{1}{2\sqrt{z}}\right) & z>\frac{1}{4}\\[1ex] 
\frac{1}{2}\sqrt{1-4z}\left(\log\frac{\eta_+}{\eta_-}-i\pi\right) & z<\frac{1}{4} 
\end{array} 
\right. 
\end{array} 
\end{equation} 
and 
$\eta_\pm=\frac{1}{2}(1\pm\sqrt{1-4z})$, $a=m^2_\rho/m^2_{\pi^+}$ and  
$b=m^2_{\pi^0\pi^0}/m^2_{\pi^+}$. 
The coupling constant $g$ comes from the strong amplitude 
${\cal A}(\rho\rightarrow\pi^+\pi^-)=-\sqrt{2}g\,\epsilon^\ast\cdot (p_+-p_-)$  
with $|g|=4.27$ to agree with $\Gamma (\rho\rightarrow\pi^+\pi^-)_{\rm exp}= 150.2$ MeV. 
The latter is the part beyond standard ChPT which we have fixed phenomenologically. 
The four-pseudoscalar amplitude is instead a standard ChPT amplitude 
which is found to depend linearly on the variable $s=m^2_{\pi^0\pi^0}$: 
\begin{equation} 
\label{A4PChPTphys} 
{\cal A}(\pi^+\pi^-\rightarrow\pi^0\pi^0)_\chi=\frac{s-m^2_\pi}{f_\pi^2}\ . 
\end{equation} 
Notice that this ChPT amplitude factorizes in Eq.~(\ref{ArhoChPT}).  
 
The invariant mass distribution for the $\rho\rightarrow\pi^0\pi^0\gamma$ decay  
is predicted to be\footnote{In terms of the photon energy, 
$E_\gamma=(m^2_\rho-m^2_{\pi^0\pi^0})/(2m_\rho)$, the photonic spectrum is written as 
$d\Gamma/dE_\gamma=(m_\rho/m_{\pi^0\pi^0})\times d\Gamma/dm_{\pi^0\pi^0}$.}: 
\begin{equation} 
\label{dGChPT} 
\begin{array}{rl} 
\frac{d\Gamma(\rho\rightarrow\pi^0\pi^0\gamma)_\chi}{dm_{\pi^0\pi^0}}&=\  
\frac{\alpha}{192\pi^5}\frac{g^2}{4\pi}\frac{m^4_\rho}{m^4_{\pi^+}} 
\frac{m_{\pi^0\pi^0}}{m_\rho}\left(1-\frac{m^2_{\pi^0\pi^0}}{m^2_\rho}\right)^3 
\sqrt{1-\frac{4m^2_{\pi^0}}{m^2_{\pi^0\pi^0}}}\\[2ex] 
&\times\ |L(m^2_{\pi^0\pi^0})|^2 
|{\cal A}(\pi^+\pi^-\rightarrow\pi^0\pi^0)_\chi|^2\ . 
\end{array} 
\end{equation} 
Integrating Eq.~(\ref{dGChPT}) over the whole physical region one obtains  
$\Gamma(\rho\rightarrow\pi^0\pi^0\gamma)_{\chi}$ =1.55 keV and  
\begin{equation} 
\label{BRChPT} 
B(\rho\rightarrow\pi^0\pi^0\gamma)_\chi=1.0\times 10^{-5}\ . 
\end{equation} 
These results confirm and update the prediction for this process given in  
Ref.~\cite{Bramon:1992ki}\footnote{With the numerical input used in 
Ref.~\protect\cite{Bramon:1992ki} one obtains 
$\Gamma(\rho\rightarrow\pi^0\pi^0\gamma)_{\chi}=1.42$ keV.}.  
   
\section{Scalar meson exchange in $\rho\rightarrow\pi^0\pi^0\gamma$} 
\label{sectLsigmaM} 
 
We now turn to the contributions coming from scalar resonance exchange.  
From a ChPT perspective their effects are encoded in the low energy constants of the  
higher order pieces of the ChPT Lagrangian. 
But the existence of a low mass $\sigma(500)$ meson should manifest in the  
$\rho,\omega\rightarrow\pi^0\pi^0\gamma$ decays not as a constant term but rather  
through a more complex resonant amplitude.  
In this section, we propose a $\sigma(500)$ dominated $\rho\rightarrow\pi^0\pi^0\gamma$ 
amplitude which coincides with the previous ChPT amplitude in the low part of the  
$\pi^0\pi^0$ invariant mass spectrum. 
In this respect, our proposed amplitude obeys the ChPT dictates 
but it also generates the resonant $\sigma(500)$ meson effects for the higher part of the 
$\pi^0\pi^0$ spectrum. 
 
\begin{figure} 
\centerline{\includegraphics[width=0.85\textwidth]{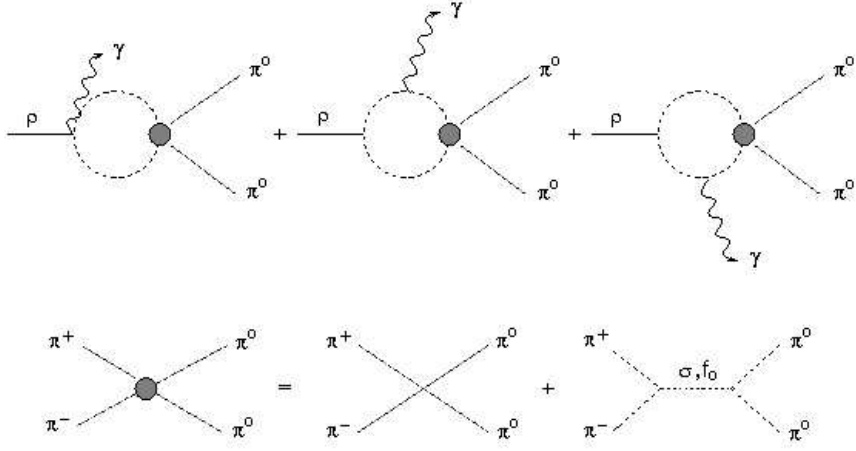}}  
\caption{\small 
One-loop Feynman diagrams for $\rho\rightarrow\pi^0\pi^0\gamma$ in the L$\sigma$M.} 
\label{figloops} 
\end{figure}

The Linear Sigma Model (L$\sigma$M) \cite{Levy:1967,Gasiorowicz:1969kn,Schechter:1971}  
will be shown to be particularly appropriate for our purposes.  
It is a well-defined $U(3)\times U(3)$ chiral model which incorporates {\it ab initio} 
both the nonet of pseudoscalar mesons together with its chiral partner,  
the scalar mesons nonet. 
In this context, the $V\rightarrow P^0 P^0\gamma$ decays proceed through a loop of  
charged pseudoscalar mesons emitted by the initial vector.  
Because of the additional emission of a photon, these charged pseudoscalar pairs with  
the initial $J^{PC}=1^{--}$ quantum numbers can rescatter into $J^{PC}=0^{++}$ pairs of 
charged or neutral pseudoscalars.   
For the $\rho\rightarrow\pi^0\pi^0\gamma$ decay the contributions from charged kaon loops 
are again negligible compared to those from pion loops and will not be considered.   
The $\sigma(500)$ and $f_0(980)$ scalar resonances are then expected to play the central 
r\^ole in this $\pi^+\pi^-\rightarrow\pi^0\pi^0$ rescattering process  
(see Fig.~\ref{figloops}) 
and the L$\sigma$M seems mostly appropriate to fix the corresponding amplitudes.  

A straightforward calculation of the $\rho\rightarrow\pi^0\pi^0\gamma$ decay amplitude  
leads to an expression identical to that in Eq.~(\ref{ArhoChPT}) but with the  
four-pseudoscalar amplitude now computed in a L$\sigma$M context, {\it i.e.} 
\begin{equation} 
\label{A4PLsigmaM} 
{\cal A}(\pi^+\pi^-\rightarrow\pi^0\pi^0)_{\mbox{\scriptsize L$\sigma$M}}= 
g_{\pi^+\pi^-\pi^0\pi^0}-\frac{g_{\sigma\pi^+\pi^-}g_{\sigma\pi^0\pi^0}}{D_{\sigma}(s)} 
                        -\frac{g_{f_0\pi^+\pi^-}g_{f_0\pi^0\pi^0}}{D_{f_0}(s)}\ , 
\end{equation} 
where $D_{S}(s)=s-m^2_S +i\,m_S\Gamma_S$ are the $S=\sigma, f_0$ propagators. 
The various coupling constants are fixed within the model and can be expressed in terms of  
$f_{\pi}$, the masses of the pseudoscalar and scalar mesons involved in the process, 
and the scalar meson mixing angle in the flavour basis  
$\phi_S$ \cite{Napsuciale:1998ip,Tornqvist:1999tn,Delbourgo:1999gi}.   
This amplitude can then be rewritten as 
\begin{equation} 
\label{A4PLsigmaMphys} 
{\cal A}(\pi^+\pi^-\rightarrow\pi^0\pi^0)_{\mbox{\scriptsize L$\sigma$M}}=   
\frac{s-m^2_\pi}{f_\pi^2}\times  
\left(\frac{m^2_\pi-m^2_\sigma}{D_{\sigma}(s)}{\rm c}^2\phi_S+ 
      \frac{m^2_\pi-m^2_{f_0}}{D_{f_0}(s)}{\rm s}^2\phi_S\right)\ , 
\end{equation} 
with $({\rm c}\phi_S, {\rm s}\phi_S)\equiv (\cos\phi_S, \sin\phi_S)$ respectively. 
 
A few remarks on the four-pseudoscalar amplitudes in  
Eqs.~(\ref{A4PLsigmaM},\ref{A4PLsigmaMphys}) and on their comparison with the  
ChPT amplitude in Eq.~(\ref{A4PChPTphys}) are of interest: 
 
\begin{itemize} 
\item[i)] for $m_S\rightarrow\infty$ ($S=\sigma, f_0$), the L$\sigma$M amplitude  
(\ref{A4PLsigmaMphys}) reduces to the ChPT amplitude (\ref{A4PChPTphys}). 
The former consists of a constant four-pseudoscalar vertex plus two terms whose  
$s$ dependence is generated by the scalar propagators $D_S(s)$,  
as shown in Eq.~(\ref{A4PLsigmaM}).  
Their sum (see Eq.~(\ref{A4PLsigmaMphys})) in the $m_S\rightarrow\infty$  
limit ends up with an amplitude which is linear in $s$ and mimics perfectly  
the effects of the derivative and massive terms in the ChPT Lagrangian (\ref{ChPTlag})  
leading  respectively to the two terms in the ChPT amplitude (\ref{A4PChPTphys}).  
This corresponds to the aforementioned complementarity between ChPT and  
the L$\sigma$M, and, we believe, is the main virtue of our approach making the 
whole analysis quite reliable. 
 
\item[ii)] the large widths of the scalar resonances break chiral symmetry  
if they are naively introduced in Eq.~(\ref{A4PLsigmaM}),  
an effect already noticed in Ref.~\cite{Achasov:1994iu}. 
Accordingly, we introduce the $\sigma(500)$ and $f_0(980)$ widths in the propagators 
only {\it after} chiral cancellation of constant terms in the amplitude. 
In this way the pseudo-Goldstone nature of pions is preserved. 
 
\item[iii)] the $\pi^0\pi^0$ invariant mass spectra for the  
$\rho,\omega\rightarrow\pi^0\pi^0\gamma$ decays cover the region where the presence 
of a $\sigma(500)$ meson should manifest. 
This fact makes crucial the incorporation of the $\sigma(500)$ resonance 
in an explicit way. 
The effects of the $f_0(980)$ meson, 
being its mass far from the kinematically allowed region, are expected to be negligible. 
Because of the presence of the $\sigma$ propagator, the amplitude in  
Eq.~(\ref{A4PLsigmaMphys}) 
---closely linked to that from ChPT and thus expected to account for the lowest  
   part of the $\pi^0\pi^0$ spectra--- 
should also be able to reproduce the effects of the $\sigma(500)$ pole at higher  
$\pi^0\pi^0$ invariant mass values. 
\end{itemize} 
 
In the propagators of the scalar mesons we include their total widths  
which, in principle, are predicted within the model as    
\begin{equation} 
\label{sigmawidth} 
\Gamma_{\sigma}=\frac{3m^3_{\sigma}}{32\pi f^2_\pi} 
\left(1-\frac{m^2_\pi}{m^2_\sigma}\right)^2\cos^2\phi_S 
\sqrt{1-\frac{4m^2_\pi}{m^2_\sigma}}\ , 
\end{equation} 
and a similar expression for $\Gamma_{f_0}$. 
We could also take $\phi_S\simeq -9^\circ$ which reproduces the photonic spectrum  
in $\phi\rightarrow\pi^0\pi^0\gamma$ decays where kaon loops give the most  
important contribution \cite{Lucio:1999bb}. 
However, our results are quite insensitive to the precise value of $\phi_S$ 
provided it is not too large  
(as confirmed by independent analyses \cite{Tornqvist:1999tn,Delbourgo:1999gi})  
thus making that the $\sigma(500)$ meson effects dominate over those from the 
higher mass $f_0(980)$ weakly coupled to pion pairs.  
We thus fix $\phi_S=0^\circ$ and, in this way, the relevant parameter in the  
calculation turns out to be the sigma meson mass $m_{\sigma}$. 
For its total width, $\Gamma_{\sigma}$, one can take the values predicted by  
Eq.~(\ref{sigmawidth}) as a first approximation but it seems safer to study the 
invariant mass distribution and branching ratio of $\rho\rightarrow\pi^0\pi^0\gamma$  
as a function of both parameters $m_{\sigma}$ and $\Gamma_{\sigma}$. 
Comparison with data could hopefully help to fix their values and contribute to  
decide on the existence or not of the $\sigma$ resonance.  
 
Integrating the $\pi^0\pi^0$ invariant mass spectrum for the central values of  
$m_\sigma=478^{+24}_{-23}\pm 17$ MeV and $\Gamma_{\sigma}=324^{+42}_{-40}\pm 21$ MeV, 
as recently measured by the E791 Collaboration \cite{Aitala:2001xu}, leads to  
$\Gamma(\rho\rightarrow\pi^0\pi^0\gamma)_{\mbox{\scriptsize L$\sigma$M}}=2.25$ keV  
and to the branching ratio 
\begin{equation} 
\label{BRLsigmaM} 
B(\rho\rightarrow\pi^0\pi^0\gamma)_{\mbox{\scriptsize L$\sigma$M}}=1.5\times 10^{-5}\ , 
\end{equation} 
well above the chiral loop prediction (\ref{BRChPT}).  
Similarly, for $m_\sigma=478$ MeV and a narrower width $\Gamma_\sigma=263$ MeV,  
as required by Eq.~(\ref{sigmawidth}), one predicts the larger value  
$B(\rho\rightarrow\pi^0\pi^0\gamma)_{\mbox{\scriptsize L$\sigma$M}}=2.1\times 10^{-5}$.   
Conversely, for the CLEO values $m_\sigma=555$ MeV and a much broader  
$\Gamma_\sigma=540$ MeV \cite{Asner:2000kj}, one obtains   
$B(\rho\rightarrow\pi^0\pi^0\gamma)_{\mbox{\scriptsize L$\sigma$M}}=8.3\times 10^{-6}$,  
below the chiral loop result (\ref{BRChPT}).  
These various predictions show that the branching ratio   
$B(\rho\rightarrow\pi^0\pi^0\gamma)$ is sensitive enough to the $\sigma$ meson mass and width 
to be used to extract information on these parameters.   
 
\section{Vector meson exchange in $\rho\rightarrow\pi^0\pi^0\gamma$} 
\label{sectVMD} 
 
In addition to the just discussed L$\sigma$M contributions, 
which can be viewed as an improved version  of the chiral loop predictions now  
extended to include the scalar resonance effects in a explicit way, 
$\rho,\omega\rightarrow\pi^0\pi^0\gamma$ can also proceed through  
vector meson exchange in the $t$- and $u$-channel. 
Their effects were already considered in Ref.~\cite{Bramon:1992kr} 
in a Vector Meson Dominance (VMD) context. 
In this framework $\rho\rightarrow\pi^0\pi^0\gamma$ proceeds through the exchange of 
an intermediate $\omega$ meson\footnote{$\phi$ exchange  
involves two OZI rule suppressed vertices and is totally negligible.},  
$\rho\rightarrow\omega\pi^0\rightarrow\pi^0\pi^0\gamma$, 
while $\omega\rightarrow\pi^0\pi^0\gamma$ proceeds by $\rho$ exchange. 
 
In order to describe these vector meson contributions we use the $SU(3)$  
symmetric Lagrangians 
\begin{equation} 
\label{VMDLag} 
\begin{array}{rl} 
{\cal L}_{\rm VVP}&=\ \frac{G}{\sqrt{2}}\epsilon^{\mu\nu\alpha\beta} 
                   \langle\partial_\mu V_\nu\partial_\alpha V_\beta P\rangle\ ,\\[2ex] 
{\cal L}_{{\rm V}\gamma}&=\ -4f^2 e g A_\mu\langle Q V^\mu\rangle\ , 
\end{array} 
\end{equation} 
where $G=\frac{3g^2}{4\pi^2 f}$ is the $\omega\rho\pi$ 
coupling constant \cite{Bramon:1992kr,Lucio-Martinez:2000ea}. 
The VMD amplitude for 
$\rho(q^\ast,\epsilon^\ast)\rightarrow\pi^0(p)\pi^0(p^\prime)\gamma(q,\epsilon)$ 
is then found to be   
\begin{equation} 
\label{ArhoVMD} 
{\cal A}(\rho\rightarrow\pi^0\pi^0\gamma)_{\rm VMD}=\frac{G^2 e}{\sqrt{2}g} 
\left(\frac{P^2\{a\}+\{b(P)\}}{M^2_\omega-P^2-i M_\omega\Gamma_\omega}+ 
      \frac{{P^\prime}^2\{a\}+\{b({P^\prime})\}} 
           {M^2_\omega-{P^\prime}^2-i M_\omega\Gamma_\omega}\right)\ , 
\end{equation} 
with $\{a\}$ the same as in Eq.~(\ref{ArhoChPT}) and 
\begin{equation} 
\label{b} 
\begin{array}{rl} 
\{b(P)\}&=\ -(\epsilon^\ast\cdot\epsilon)\,(q^\ast\cdot P)\,(q\cdot P) 
            -(\epsilon^\ast\cdot P)\,(\epsilon\cdot P)\,(q^\ast\cdot q)\\[1ex] 
        &\ \ \ \ +(\epsilon^\ast\cdot q)\,(\epsilon\cdot P)\,(q^\ast\cdot P) 
                 +(\epsilon\cdot q^\ast)\,(\epsilon^\ast\cdot P)\,(q\cdot P)\ , 
\end{array} 
\end{equation} 
where $P=p+q$ and $P^\prime=p^\prime+q$ are the momenta of the intermediate  
$\omega$ meson in the $t$- and $u$-channel respectively. 
From this VMD amplitude one easily obtains  
$\Gamma(\rho\rightarrow\pi^0\pi^0\gamma)_{\rm VMD}=1.88$ keV and  
\begin{equation} 
\label{BRVMD} 
B(\rho\rightarrow\pi^0\pi^0\gamma)_{\rm VMD}=1.3\times 10^{-5}\ , 
\end{equation} 
in agreement with the results in Ref.~\cite{Bramon:1992kr} once the numerical inputs  
are unified. 
\begin{figure}[t] 
\centerline{\includegraphics[width=0.85\textwidth]{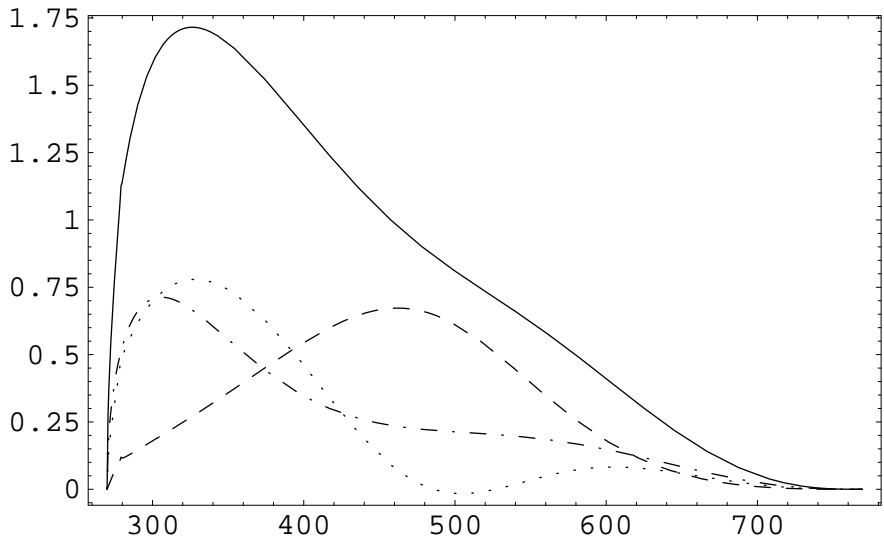}}  
\caption{\small 
$dB(\rho\rightarrow\pi^0\pi^0\gamma)/dm_{\pi^0\pi^0}\times 10^7\ (\mbox{MeV}^{-1})$ 
as a function of the dipion invariant mass $m_{\pi^0\pi^0}$ (MeV). 
The dot-dashed, dashed and dotted lines correspond to the separate contributions  
from VMD, L$\sigma$M and their interference, respectively.  
The solid line is the global result.  
The reference values $m_\sigma=478$ MeV and $\Gamma_\sigma=324$ MeV, taken from 
Ref.~\protect\cite{Aitala:2001xu}, have been used.} 
\label{figrho} 
\end{figure} 

Our final results for ${\cal A}(\rho\rightarrow\pi^0\pi^0\gamma)$  
are thus the sum of this VMD contribution plus the previously discussed L$\sigma$M  
contribution containing the scalar resonance effects. 
The corresponding $\pi^0\pi^0$ invariant mass distribution is plotted in 
Fig.~\ref{figrho}. 
The separate contributions from VMD, L$\sigma$M and their interference,  
as well as the total result are explicitly shown.  
For $m_{\sigma}$ and $\Gamma_{\sigma}$ we have taken 
$m_\sigma=478$ MeV and $\Gamma_{\sigma}=324$ MeV, 
the central values measured by the E791 Collaboration \cite{Aitala:2001xu}. 
The interference term turns out to be positive in the whole range and  
scalar meson exchange contributes decisively to increase the previous results  
as required by experiment.  
Indeed, for the integrated decay width one now obtains 
$\Gamma(\rho\rightarrow\pi^0\pi^0\gamma)_{\mbox{\scriptsize L$\sigma$M+VMD}}=5.77$ keV 
and for the branching ratio 
\begin{equation} 
\label{BRLsigmaM+VMD} 
B(\rho\rightarrow\pi^0\pi^0\gamma)_{\mbox{\scriptsize L$\sigma$M+VMD}}= 
3.8\times 10^{-5}\ . 
\end{equation} 
This value for $B(\rho\rightarrow\pi^0\pi^0\gamma)$ seems to be quite in agreement 
with the experimental result in Eq.~(\ref{SND}), 
although the current experimental error is still too big to be conclusive. 
In any case, our analysis shows the importance of including scalar resonance effects 
in an explicit way and could be taken as an indication on the existence of a  
$\sigma$ meson in the energy region around 500 MeV. 
\begin{figure}[t] 
\centerline{\includegraphics[width=0.85\textwidth]{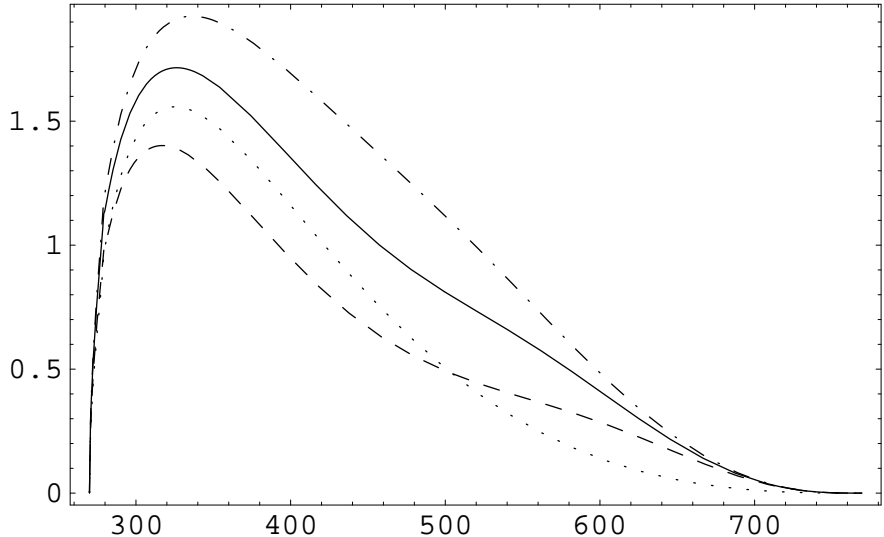}}  
\caption{\small 
$dB(\rho\rightarrow\pi^0\pi^0\gamma)/dm_{\pi^0\pi^0}\times 10^7\ (\mbox{MeV}^{-1})$ 
as a function of the dipion invariant mass $m_{\pi^0\pi^0}$ (MeV). 
The various predictions are for the input values:  
$m_\sigma = 478$ MeV and $\Gamma_\sigma = 324$ MeV    
from Ref.~\protect\cite{Aitala:2001xu} (solid line);   
$m_\sigma = 478$ MeV and $\Gamma_\sigma = 263$ MeV   
from Ref.~\protect\cite{Aitala:2001xu} and  
Eq.~(\protect\ref{sigmawidth}) (dot-dashed line); and  
$m_\sigma = 555$ MeV and $\Gamma_\sigma = 540$ MeV  
from Ref.~\cite{Asner:2000kj} (dashed line).  
The chiral loop prediction with no scalars is also included for comparison (dotted line).} 
\label{figrhoall} 
\end{figure} 

In order to show the sensitivity of our treatment on the parameters of the  
$\sigma$ meson we have plotted in Fig.~\ref{figrhoall} our final predictions for the  
$\pi^0\pi^0$ invariant mass distribution of $\rho\rightarrow\pi^0\pi^0\gamma$ for  
various values of $m_\sigma$ and $\Gamma_\sigma$. 
The shapes of the various curves are quite similar but the corresponding  
integrated values are considerably different.  
Taking now the central value of $m_\sigma=478$ MeV \cite{Aitala:2001xu} 
and $\Gamma_{\sigma}=263$ MeV, as required by Eq.~(\ref{sigmawidth}), 
one finds   
$B(\rho\rightarrow\pi^0\pi^0\gamma)_{\mbox{\scriptsize L$\sigma$M+VMD}}= 
4.7\times 10^{-5}$. 
Thus, a narrower $\Gamma_\sigma$ increases $B(\rho\rightarrow\pi^0\pi^0\gamma)$ 
to a value which almost coincides with the central value of the  
SND measurement (\ref{SND}).  
The prediction for the values $m_\sigma=555$ MeV and $\Gamma_{\sigma}=540$ MeV 
reported by the CLEO Collaboration \cite{Asner:2000kj} is also included in  
Fig.~\ref{figrhoall}, 
as well as the invariant mass distribution predicted by Eq.~(\ref{dGChPT}), 
which just includes chiral loops but no scalar exchange. 
In these cases the corresponding branching ratios are found to be  
$2.8\times 10^{-5}$ and $2.9\times 10^{-5}$, respectively, 
well below the SND data in Eq.~(\ref{SND}). 
The smallness of the former value disfavours a broad $\Gamma_{\sigma}$. 
The second value is an update of the old result in Ref.~\cite{Bramon:1992ki}  
and its smallness confirms the need of the effects of a narrow ${\sigma}$.  
 
\section{$\omega\rightarrow\pi^0\pi^0\gamma$} 
\label{sectomega} 
 
The $\omega\rightarrow\pi^0\pi^0\gamma$ radiative decay can now be treated along 
the same lines. 
This process receives a well known $\rho$ meson exchange contribution 
via the VMD decay chain  
$\omega\rightarrow\rho\pi^0\rightarrow\pi^0\pi^0\gamma$ \cite{Bramon:1992kr}.   
Ignoring for the moment $\rho$-$\omega$ mixing, 
{\it i.e.}~assuming that the physical $\omega =\omega^{I=0}$ with no $I=1$ contaminations, 
the corresponding amplitude is given by 
${\cal A}^{I=0}(\omega\rightarrow\pi^0\pi^0\gamma)_{\rm VMD}=\frac{1}{3} 
 {\cal A}(\rho\rightarrow\pi^0\pi^0\gamma)_{\rm VMD}$ 
with the replacement $(M_\rho, \Gamma_\rho)\rightarrow (M_\omega, \Gamma_\omega)$ 
in the propagators of Eq.~(\ref{ArhoVMD}). 
The proportionality factor $1/3$ follows from the 
$SU(3)$ symmetric Lagrangians (\ref{VMDLag}) and for an ideally mixed $\omega$.  
Since the $\pi^0\gamma$ invariant masses are far from the $\rho$ poles, 
this amplitude is nearly real as before and the invariant $\pi^0\pi^0$ mass distribution 
has a similar shape to that of the $\rho\rightarrow\pi^0\pi^0\gamma$ case. 
Integrating over the whole physical region one obtains  
$\Gamma(\omega\rightarrow\pi^0\pi^0\gamma)_{\rm VMD}=268\ \mbox{eV}$   
and  
$B(\omega\rightarrow\pi^0\pi^0\gamma)_{\rm VMD}=3.2\times 10^{-5}$,  
in agreement with the results of Ref.~\cite{Bramon:1992kr}.  
If instead we use a momentum dependent width for the $\rho$ meson \cite{O'Connell:1997wf} 
\begin{equation} 
\label{rhowidth} 
\Gamma_{\rho}(q^2)=\Gamma_{\rho} 
\left(\frac{q^2-4m^2_\pi}{m^2_\rho -4m^2_\pi}\right)^{3/2}\frac{m_\rho}{\sqrt{q^2}}\, 
\theta (q^2-4m^2_\pi)\ , 
\end{equation} 
then one obtains  
$\Gamma(\omega\rightarrow\pi^0\pi^0\gamma)_{\rm VMD}=300$ eV.   
This value is some 12\% larger than the previous one, as already noticed in 
Ref.~\cite{Guetta:2001ra}.  
Notice that our results are still substantially lower than the central value 
reported in Ref.~\cite{Guetta:2001ra}.  
The reason is that we are using an $SU(3)$ symmetric formalism where all 
the $VVP$ and $VP\gamma$ couplings are deduced from the $VPP$ coupling $g$, 
which we take from the $\rho\rightarrow\pi^+\pi^-$ width 
(see Ref.~\cite{daphne95:bramon} for details), 
while in Ref.~\cite{Guetta:2001ra} the couplings  
$g_{\omega\rho\pi}$ and $g_{\rho^0\pi^0\gamma}$ are extracted from experiment. 
In principle, this seems a better procedure but, unfortunately,  
the extraction of $g_{\omega\rho\pi}$ from  
$\Gamma(\omega\rightarrow\pi^+\pi^-\pi^0 )_{\rm exp}$ is based on the assumption that  
this decay proceeds entirely through 
$\omega\rightarrow\rho\pi\rightarrow\pi^+\pi^-\pi^0$ 
and $g_{\rho^0\pi^0\gamma}$ follows from the experimental value  
$\Gamma(\rho^0\rightarrow\pi^0\gamma )_{\rm exp}=(102\pm 26)\ \mbox{keV}$ 
which is controversial and affected by large errors. 
If we use this value, our predictions increase by some 19\% and confirm the result 
$\Gamma(\omega\rightarrow\pi^0\pi^0\gamma)_{\rm VMD}=(344\pm 85)$ eV 
of Ref.~\cite{Guetta:2001ra}. 
 
There is also another contribution to the $\omega\rightarrow\pi^0\pi^0\gamma$ amplitude 
coming from chiral loops. 
However, as stated in the Introduction, this chiral loop contribution  
(given only by kaon loops in the good isospin limit with $\omega =\omega^{I=0}$) 
is very small and can be safely neglected.   
Its improved version taking into account scalar resonance effects is more problematic 
because kaons could couple to the $\sigma$ meson. 
Proceeding as before one can obtain the 
${\cal A}(K^+K^-\rightarrow\pi^0\pi^0)_{\mbox{\scriptsize L$\sigma$M}}$ amplitude 
corresponding to those in 
Eqs.~(\ref{A4PLsigmaM},\ref{A4PLsigmaMphys}). 
There is however an important difference: 
while the $g_{\sigma\pi\pi}$ couplings are proportional to $(m^2_{\sigma}-m^2_\pi)$, 
those for $g_{\sigma K\bar K}$ are proportional to $(m^2_{\sigma}-m^2_K)$. 
For the range of masses we are considering, $m_{\sigma}\simeq m_K$, 
the amplitude containing the $\sigma$ pole turns out to be negligible. 
In this case we still have  
${\cal A}^{I=0}(\omega\rightarrow\pi^0\pi^0\gamma)\simeq 
 {\cal A}^{I=0}(\omega\rightarrow\pi^0\pi^0\gamma)_{\rm VMD}$ 
as emphasized in Ref.~\cite{Guetta:2001ra}. 
From these various estimates, reflecting the large uncertainties in this channel,  
it seems reasonable to conclude   
\begin{equation} 
\label{GomegaVMD} 
\Gamma(\omega\rightarrow\pi^0\pi^0\gamma)_{\rm VMD}=(330\pm 90)\ \mbox{eV}\ , 
\end{equation} 
quite close to the value favoured in Ref.~\cite{Guetta:2001ra} and  
affected by a conservative error.  
  
In addition to the dominant VMD contribution there is an indirect contribution to 
$\omega\rightarrow\pi^0\pi^0\gamma$ that appears through $\rho$-$\omega$ mixing 
followed by the $\rho\rightarrow\pi^0\pi^0\gamma$ decay \cite{Guetta:2001ra}. 
This new contribution makes the whole $\omega\rightarrow\pi^0\pi^0\gamma$ amplitude 
to be written as  
${\cal A}^{I=0}(\omega\rightarrow\pi^0\pi^0\gamma)+\epsilon 
 {\cal A}(\rho\rightarrow\pi^0\pi^0\gamma)$, 
with two amplitudes already discussed and where $\epsilon$ is the 
$\rho$-$\omega$ mixing parameter given by 
\begin{equation} 
\label{epsilon} 
\epsilon \equiv \frac{{\cal M}^2_{\rho\omega}} 
              {m^2_{\omega}-m^2_{\rho}-i(m_\omega\Gamma_{\omega}-m_\rho\Gamma_{\rho})} 
\simeq -0.006+i\,0.034\ , 
\end{equation} 
with  
${\cal M}^2_{\rho\omega}(m^2_\rho)=(-3800\pm 370)\ \mbox{MeV}^2$ \cite{O'Connell:1997wf}. 
An additional effect of this $\rho$-$\omega$ mixing is to replace the $\rho$ propagator  
in ${\cal A}^{I=0}$ by 
\begin{equation} 
\label{Drho} 
\frac{1}{D_\rho(s)}\rightarrow\frac{1}{D^\epsilon_\rho(s)}=\frac{1}{D_\rho(s)} 
\left(1+\frac{g_{\omega\pi\gamma}}{g_{\rho\pi\gamma}} 
        \frac{{\cal M}^2_{\rho\omega}}{D_\omega(s)}\right)\ , 
\end{equation} 
with $D_V(s)=s-m^2_V+i\,m_V\Gamma_V$ for $V=\rho, \omega$ and in our $SU(3)$ symmetric  
VMD framework $g_{\omega\pi\gamma}/g_{\rho\pi\gamma}=3$. 
\begin{figure}[t] 
\centerline{\includegraphics[width=0.85\textwidth]{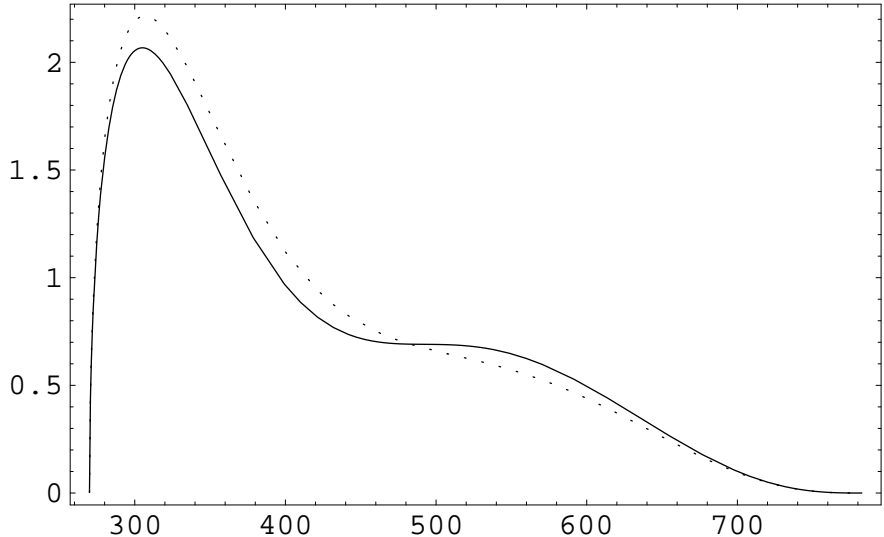}}  
\caption{\small 
$dB(\omega\rightarrow\pi^0\pi^0\gamma)/dm_{\pi^0\pi^0}\times 10^7\ (\mbox{MeV}^{-1})$ 
as a function of the dipion invariant mass $m_{\pi^0\pi^0}$ (MeV). 
The predictions are for the $\sigma$ meson values 
$m_\sigma=478$ MeV and $\Gamma_\sigma=324$ MeV (solid line) and 
dropping all $\sigma$ meson contribution (dotted line)} 
\label{figomega} 
\end{figure} 

Apparently, the authors of Ref.~\cite{Guetta:2001ra} have approximated the new,  
isospin violating term of $\omega\rightarrow\pi^0\pi^0\gamma$ by the VMD contribution 
$\epsilon {\cal A}(\rho\rightarrow\pi^0\pi^0\gamma)_{\rm VMD}$. 
In so doing one increases the previous estimate to   
$\Gamma(\omega\rightarrow\pi^0\pi^0\gamma)=(381\pm 90)$ eV 
quite close to the result in Ref.~\cite{Guetta:2001ra}. 
A more complete treatment, with  
${\cal A}(\omega\rightarrow\pi^0\pi^0\gamma)=  
 {\cal A}^{I=0}(\omega\rightarrow\pi^0\pi^0\gamma)+\epsilon 
 {\cal A}(\rho\rightarrow\pi^0\pi^0\gamma)_{\mbox{\scriptsize VMD+L$\sigma$M}}$, 
seems however preferable. 
The $\pi^0\pi^0$ invariant mass spectra corresponding to this amplitude  
have been calculated for the same input values of $m_\sigma$ and $\Gamma_\sigma$ 
that we introduced in the $\rho\rightarrow\pi^0\pi^0\gamma$ case.  
But the sensitivity on these input parameters is now minimal and all the results 
almost coincide with the curve for $m_\sigma=478$ MeV and $\Gamma_\sigma=324$ MeV 
\cite{Aitala:2001xu} plotted in Fig.~\ref{figomega}. 
 
The integrated width and branching ratio are predicted to be  
$\Gamma(\omega\rightarrow\pi^0\pi^0\gamma)_{\mbox{\scriptsize VMD+L$\sigma$M}}= 
 (377\pm 90)$ eV and  
\begin{equation} 
\label{Bomegatotal} 
B(\omega\rightarrow\pi^0\pi^0\gamma)_{\mbox{\scriptsize VMD+L$\sigma$M}}= 
(4.5\pm 1.1)\times 10^{-5}\ . 
\end{equation} 
If the chiral loops are retained but scalar meson effects are neglected 
one then predicts  
$\Gamma(\omega\rightarrow\pi^0\pi^0\gamma)_{\mbox{\scriptsize VMD+$\chi$}}= 
 (395\pm 90)$ eV and  
\begin{equation} 
\label{Bomegatotalch} 
B(\omega\rightarrow\pi^0\pi^0\gamma)_{\mbox{\scriptsize VMD+$\chi$}}= 
(4.7\pm 1.1)\times 10^{-5}\ ,  
\end{equation} 
only a 5\% above the previous results and hardly distinguishable. 
The same happens to the invariant mass distribution also plotted in Fig.~\ref{figomega}. 
Because of the large errors, the agreement with the experimental measurement (\ref{GAMS}) 
is reasonable but a moderate improvement of the data will represent a decisive test for  
our approach.  
  
\section{Conclusions} 
 
In this note we have discussed scalar and vector meson exchange in  
$\rho,\omega\rightarrow\pi^0\pi^0\gamma$ decays.   
Vector meson contributions are calculated in the framework of VMD and confirm the  
old results in Ref.~\cite{Bramon:1992kr}. 
The scalar meson contributions are much more interesting and have been introduced by  
means of a ChPT inspired context first applied to $\phi\rightarrow\pi^0\eta\gamma$ 
\cite{Bramon:2000vu}. 
The main point in this context is the use of an amplitude which agrees with ChPT for 
low values of the two-pseudoscalar invariant mass but develops the scalar meson poles 
at higher values in accordance with the L$\sigma$M Lagrangian.   
 
Besides a sizeable VMD contribution to $\rho\rightarrow\pi^0\pi^0\gamma$,   
there also exists a larger contribution coming from pion loops which couple strongly to the 
low mass $\sigma$ meson.  
The predictions for the $\pi^0 \pi^0$ invariant mass distribution and the integrated 
$\rho\rightarrow\pi^0\pi^0\gamma$ width are sensitive enough  
to $m_\sigma$ and $\Gamma_\sigma$ to allow for interesting 
comparisons with experiment. 
The recently available data for $B(\rho\rightarrow\pi^0\pi^0\gamma)$ in Eq.~(\ref{SND}) 
from the SND Collaboration favour the presence of a low mass and moderately narrow  
$\sigma$ meson.  
 
The parallel analysis of the $\omega\rightarrow\pi^0\pi^0\gamma $ decay is more involved 
because $\rho$-$\omega$ mixing plays a r\^ole, 
as first analyzed by Guetta and Singer \cite{Guetta:2001ra}. Moreover, for this decay  
the main contribution comes from a less well fixed VMD amplitude and  
the effects of scalar meson exchange are much more difficult to disentangle. In this case,   
there is little hope to learn on the values of $m_\sigma$ and $\Gamma_\sigma$ 
when comparing with experiment. 
The available data in Eq.~(\ref{GAMS}) are compatible with our predictions, 
although poorly conclusive because they are affected by large errors.  
 
In summary,  
higher accuracy data for these two channels and more refined theoretical analyses    
would contribute decisively to clarify one of the challenging aspects of present  
hadron physics, namely, the  structure of the lowest lying scalar states and  
particularly of the controversial $\sigma$ meson.  
 
\section*{Acknowledgements} 
The authors acknowledge enlightening discussions with M.~D.~Scadron.  
Work partly supported by the EEC, TMR-CT98-0169, EURODAPHNE network. 
M.~N.~was supported by CONCYTEG-Mexico under project 00-16-CONCY\-TEG/CONACYT-075. 
J.~L.~L.~M.~acknowledges the warm hospitality from IFAE and partial financial support 
from IFAE, CONACyT and CONCyTEG.

\end{document}